\documentclass[prl,twocolumn,showpacs,preprintnumbers,amsmath,amssymb]{revtex4}

\usepackage[dvips]{graphicx}
\usepackage{dcolumn}
\usepackage{bm}

\begin{document}

\preprint{}

\title{Spontaneous Symmetry Breaking in Superfluid Helium-4}

\author{Junpei Harada}
 \email[]{E-mail:jharada@apctp.org}
 \affiliation{Asia Pacific Center for Theoretical Physics, Pohang,
 790-784, Korea \\
 Physics division, History of Science,
 Kyoto University, Kyoto, 606-8501, Japan}%
\date{January 16, 2007}

\begin{abstract}
We derive an analytical expression for a critical temperature of
 spontaneous symmetry breaking in a repulsive hard-core interacting Bose system.
We show that the critical temperature of spontaneous symmetry breaking
 in a hard-core interacting Bose system is determined
 by the three physical parameters: the density of Bose liquid at absolute
 zero ($\rho_0$), the mass ($m$) and the hard sphere diameter ($\sigma$)
 of a boson.
The formula that we have derived is $T_c = \rho_0 \pi \hbar^2 \sigma/m^2 k_B$.
We report $T_c$ of liquid helium-4 is
 2.194 K, which is significantly close to the $\lambda$-temperature of
 2.1768 K.
The deviation between the predicted and experimental values of the
 $\lambda$-temperature is less than 1$\%$.
\end{abstract}


\maketitle
\noindent{\bf 1.}
The year 2008 is the 100th anniversary of ``Liquid Helium Year'', in
which Heike Kamerlingh Onnes produced liquid helium.
We are in a truly memorial moment in the history of low temperature physics.
Liquid helium has been studied for the past 100 years, and
there is no question that liquid helium is of central importance in this field.
(I would like to turn readers' attention to Ref.~\cite{Laesecke:2002}.
It presents the first complete English translation of the inaugural speech of
Heike Kamerlingh Onnes at the University of Leiden in 1882.
Although his speech is not related to helium, 
it is quite interesting and gives us a lesson.
I believe that it attracts the interest of readers.)

In spite of great efforts, an important problem still remains to be solved.
The problem is as follows.
Liquid helium-4 undergoes a phase transition, known as the
$\lambda$-transition, to a superfluid phase at the $\lambda$-temperature.
The experimental value of the $\lambda$-temperature $T_\lambda$ is
approximately equal to 2.2 K.
The problem is quite simple.
Why is $T_\lambda \simeq 2.2$ K?
What physical parameters determine the value of the $\lambda$-temperature?
This is the ``Why 2.2 K?'' problem, which is the subject of the present paper.
It is a long-standing dream in low temperature physics to 
derive a formula for the liquid helium-4 $\lambda$-temperature.

In 1938, Fritz London proposed that the $\lambda$-transition of liquid
helium-4 probably has to be regarded as the Bose-Einstein
condensation (BEC)~\cite{London:1938_1}.
He calculated a critical temperature of the BEC
in an ideal Bose gas:
\begin{eqnarray}
 T_{BE} = \frac{2\pi \hbar^2}{m k_B}
          \left(\frac{n}{\zeta (3/2)}\right)^{2/3}, \label{eq:BEC}
\end{eqnarray}
where $m$ is a boson mass, $n$ is a number density of Bose gas and
 $\zeta (3/2) = 2.612375\cdots$. Thus, the Bose-Einstein temperature $T_{BE}$
 is determined by the {\it two} physical parameters, $m$ and $n$.
 London reported $T_{BE}$ of helium is 3.13 K~\cite{London:1938_2},
 which is the same order of magnitude as the $\lambda$-temperature of
 2.18 K.
His proposal is quite important, because it was the first suggestion
that the $\lambda$-transition was intrinsically related to the
Bose-Einstein condensation.
Recall that it was not clear in the 1930s whether the BEC was an actually physical phenomenon~\cite{London:1938_2}.

Although London's work was revolutionary progress, 
there is a difference of approximately 0.95 K between his
theoretical prediction and the experimental value.
This deviation is approximately 44$\%$ and it is not negligible.
It is widely believed that this discrepancy arises from the neglect of 
interactions between helium atoms in his theoretical
calculation. This conjecture is quite reasonable
because liquid helium is not an ideal Bose gas but a repulsive interacting
Bose system.
However, in general, it is difficult to consider interactions between
atoms in analytical calculations of the BEC.
Although this problem has been discussed~\cite{Williams:1995}, it
remains very important.

In this paper, we study the ``Why 2.2 K?'' problem via an
alternative approach.
The key concept in our approach is {\it spontaneous symmetry breaking}.
We derive an analytical formula for a critical temperature of
spontaneous symmetry breaking in a repulsive hard-core interacting Bose system.
We show that a critical temperature is determined by the {\it three}
physical parameters: the density
of Bose liquid at absolute zero ($\rho_0$), the mass ($m$) and the hard
sphere diameter ($\sigma$) of a boson.
Our formula predicts that a critical temperature of liquid helium-4  
is 2.194 K, which is significantly close to the experimental value of
the $\lambda$-temperature ($T_\lambda$ = 2.1768 K~\cite{Donnelly:1998}).

\noindent{\bf 2.}
Spontaneous symmetry breaking is a symmetry breaking by the ground state of a system.
The symmetry can be discrete or continuous, and be local or global.
Spontaneous symmetry breaking is a phenomenon that occurs in many systems.
In condensed matter physics, ferromagnetism is a primary example.
For superconductivity, the importance of spontaneous symmetry breaking
was firstly emphasized by Yoichiro Nambu~\cite{Nambu:1960tm}.
In particle physics, the electroweak symmetry breaking is established in
the Standard Model.
In cosmology, a number of spontaneous symmetry breaking play an
important role in the history of the early universe.
Furthermore, the recently proposed {\it ghost
condensation}~\cite{Arkani-Hamed:2003uy} in the accelerating universe
can be regarded as a kind of spontaneous symmetry breaking.
Thus, spontaneous symmetry breaking is a very important concept in modern
physics.

Spontaneous symmetry breaking is crucial in liquid helium-4. 
In short, {\it the $\lambda$-transition is the spontaneous symmetry breaking}.
This is the basis of our approach.
We derive a formula for a critical temperature of spontaneous symmetry
breaking in the framework of effective field theory described by the
order parameters.
In the case of liquid helium-4, the order parameter is
a macroscopic wavefunction $\varphi$,
which is a one-component complex scalar field with both amplitude
and phase.
We begin with the theory described by this scalar field $\varphi$.

\noindent{\bf 3.}
The Lagrangian density ${\cal L}( = {\cal K} - {\cal V})$ 
of the system is given by the nonrelativistic Goldstone model of the form
\begin{eqnarray}
{\cal V} = - \mu\varphi^*\varphi +
 \frac{\lambda}{2}(\varphi^*\varphi)^2, \qquad
 \lambda = \frac{2\pi \hbar^2 \sigma}{m}.
 \label{eq:potential}
\end{eqnarray}
The kinetic part ${\cal K} = i\hbar \varphi^* \partial_t \varphi +
\hbar^2 \varphi^* \nabla^2 \varphi/2m$ is not important here, and therefore we
concentrate on the potential ${\cal V}$.
In Eq.~(\ref{eq:potential}), $\mu$ is the chemical potential, $\lambda$
is the coupling constant for helium interactions and $\varphi^*$ is
the complex conjugate of $\varphi$.
The interaction between helium atoms is repulsive at short distances
(hard-core interactions). 
Therefore, the coupling constant $\lambda = 2\pi \hbar^2 \sigma/m $ is
positive, where $m$ and $\sigma$ are the mass and the hard sphere
diameter of a helium atom, respectively.

We first consider the physical dimensions of parameters of the
potential~(\ref{eq:potential}).
In units of $\hbar = 1$ and $k_B = 1$ ($\hbar$ is the reduced Planck's
constant and $k_B$ is the Boltzmann constant), for any 
physical quantity $Q$, its physical dimension $[Q]$ is written as the
product of the dimensions of length and temperature:
\begin{eqnarray}
 [Q] = [\mbox{Length}]^\alpha [\mbox{Temperature}]^\beta,
\end{eqnarray}
where $\alpha$ and $\beta$ are numerical constants.
The physical dimension of the Lagrangian density is $[{\cal L}] =
[\mbox{L}]^{-3} [\mbox{T}]$, and that of the scalar field $\varphi$ is
$[\varphi] = [\mbox{L}]^{-3/2}$. (Here $[\mbox{L}]$ and $[\mbox{T}]$
represent $[\mbox{Length}]$ and $[\mbox{Temperature}]$, respectively.)
Hence it is straightforward to derive the physical dimensions of
parameters in Eq.~(\ref{eq:potential}):
\begin{eqnarray}
 [\mu] = [\mbox{T}], \
 [\lambda] = [\mbox{L}]^3 [\mbox{T}], \
 [m] = [\mbox{L}]^{-2} [\mbox{T}]^{-1}, \
 [\sigma] = [\mbox{L}].
\end{eqnarray}
Note here that in the present case the coupling constant $\lambda$ is
a dimensionful parameter,
in contrast to a relativistic case in which $\lambda$ is
dimensionless.
These physical dimensions help us to understand the
physics of liquid helium-4.

We next consider the symmetry of Lagrangian.
The potential ${\cal V}$ of Eq.~(\ref{eq:potential}) is invariant under
the U(1) phase transformation of the field $\varphi$: 
$\varphi \rightarrow e^{i\theta}
\varphi$, $\varphi^* \rightarrow e^{-i\theta} \varphi^*$.
Furthermore, if this is a global transformation in which the
transformation parameter $\theta$ does not depend on the space and time
coordinates, the kinetic part ${\cal K}$ is invariant under the U(1)
phase transformation.
Hence, the system described by the Lagrangian density ${\cal L}$ has
a global U(1)$\simeq$O(2) symmetry.
This global U(1) symmetry is
spontaneously broken at low temperature as you see below.

\begin{figure}[t]
\includegraphics[width=5cm]{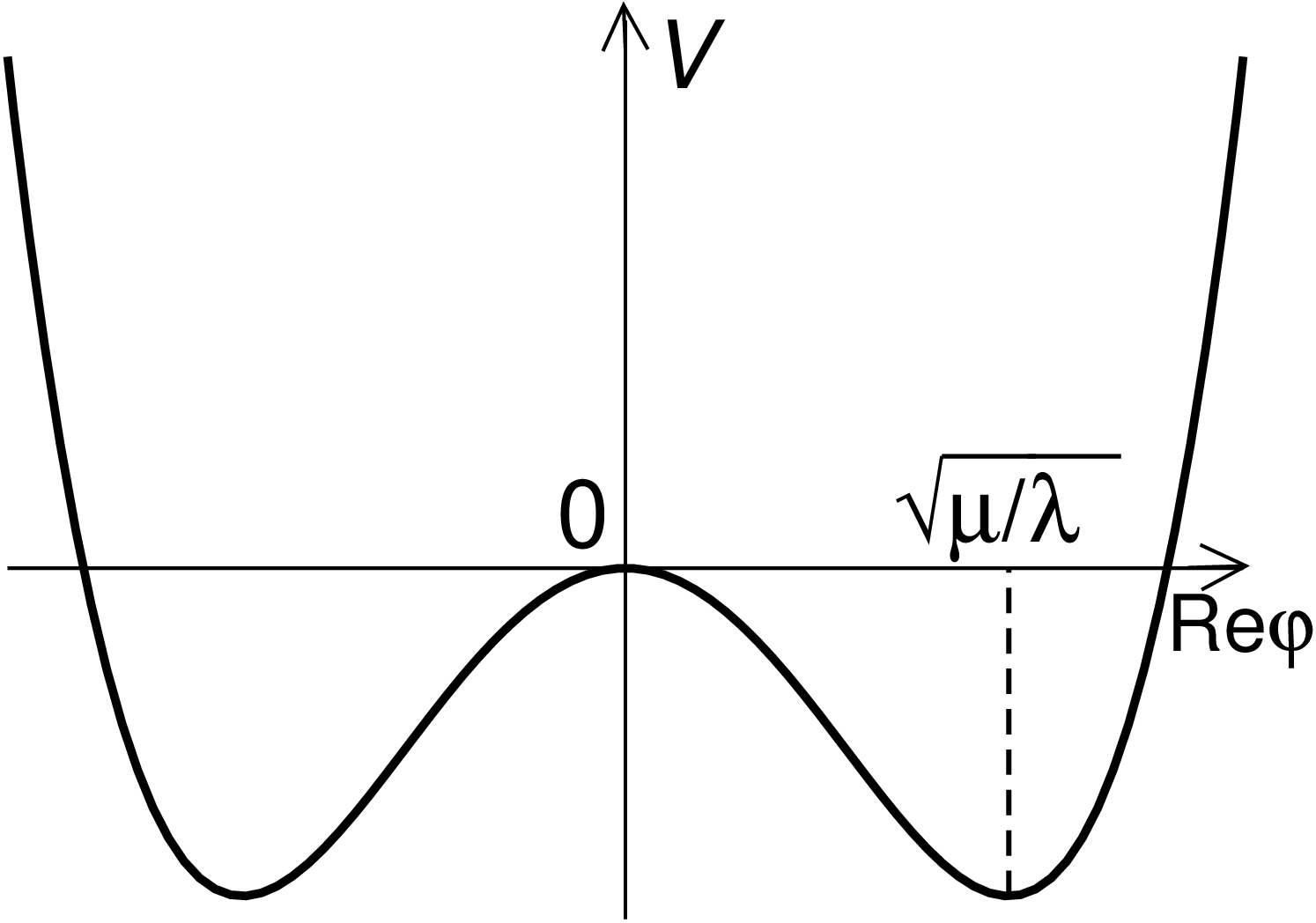}
\caption{\label{fig:potential} The potential ${\cal V}$ in the case that $\mu > 0$.}
\end{figure}

\noindent{\bf 4.}
The ground state of a system is given by solving the condition 
$\partial {\cal V}/\partial \varphi = 0$.
First, we consider the case that a chemical potential $\mu$ is {\it
negative} ($\mu \leq 0$).
In this case, there is only a trivial solution $\varphi = 0$.
This solution is the stable ground state, because
$\partial^2 {\cal V}/\partial \varphi^* \partial \varphi |_{\varphi = 0}
\geq 0$.
Hence the global U(1) symmetry is unbroken in this case.
Second, we consider the case that $\mu$ is {\it positive} ($\mu > 0$).
In this case, in contrast to the case that $\mu \leq 0$, there is a
nontrivial solution in addition to a trivial solution $\varphi = 0$:
\begin{eqnarray}
 |\varphi| = \sqrt{\frac{\mu}{\lambda}}.
\end{eqnarray}
Fig.~\ref{fig:potential} shows the potential ${\cal V}$ in the case that
$\mu > 0$.
In the present case, the trivial solution $\varphi = 0$ is unstable
 and it is not the ground state of the system.
The nontrivial solution $|\varphi| = \sqrt{\mu
/\lambda}$ is the stable ground state, because
the curvature of the potential at two solutions
satisfies
\begin{eqnarray}
 \frac{\partial^2 {\cal V}}{\partial \varphi^* \partial
  \varphi}\Big|_{\varphi = 0} < 0, \quad
 \frac{\partial^2 {\cal V}}{\partial \varphi^* \partial
  \varphi}\Big|_{|\varphi| = \sqrt{\mu /\lambda}} > 0.
\end{eqnarray}

Therefore, if a chemical potential $\mu$ is {\it positive} ($\mu > 0$),
the global U(1) symmetry is spontaneously broken.
From above arguments, we reach a following view.
A physical system undergoes a phase transition at a critical temperature $T_c$, 
if a chemical potential satisfies the conditions:
$\mu \leq 0 \ (T \geq T_c)$, $\mu > 0 \ (T < T_c)$.
Spontaneous symmetry breaking occurs only if a chemical potential
becomes {\it positive}.
Therefore, in contrast to the Bose-Einstein condensation, 
{\it spontaneous symmetry breaking never occurs in an ideal Bose gas},
in which a chemical potential is always {\it negative}
($\mu \leq 0$).
This is a crucial difference between spontaneous symmetry breaking and
the standard Bose-Einstein condensation.

\noindent{\bf 5.}
Before deriving a formula for a critical temperature, 
we consider the temperature dependence of a chemical potential $\mu$.
Although the temperature dependence of $\mu$ is not necessary for 
deriving a formula, it is useful to understand a phase transition.
For this reason, we consider the temperature dependence of
$\mu$ in the following.
The quantity $|\varphi|^2$ is equivalent to the number density $n$ of
superfluid. Therefore, the following relation is satisfied below 
the $\lambda$-temperature
\begin{eqnarray}
 |\varphi | = \sqrt{\frac{\mu}{\lambda}} = \sqrt{n} =
  \sqrt{\frac{\rho_s}{m}},
\end{eqnarray}
where $\rho_s$ is the superfluid density and $m$ is a boson mass.
From this relation, the superfluid density is given by
\begin{eqnarray}
 \rho_s = \frac{m \mu}{\lambda}. \label{eq:rela}
\end{eqnarray}
This expression indicates that the temperature dependence of $\mu$ is
obtained from that of $\rho_s$.

\begin{figure}[t]
\includegraphics[width=\linewidth]{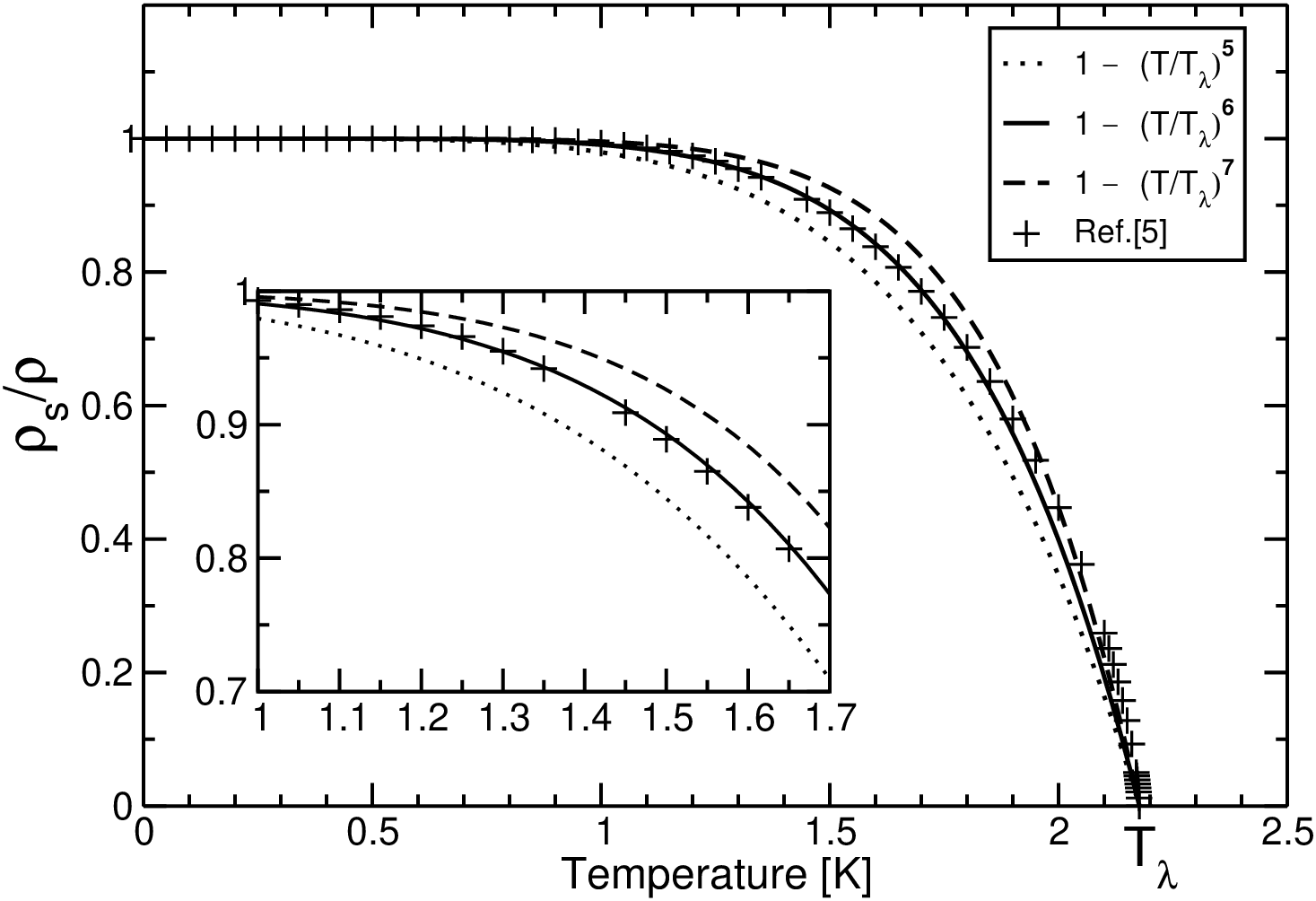}
\caption{\label{fig:ratio} Superfluid density ratio for liquid helium-4
 as a function of temperature. The inset shows the temperature region
 1.0 K $\leq T \leq$ 1.7 K. Key: (dotted line) $1-(T/T_\lambda)^5$;
 (solid line) $1-(T/T_\lambda)^6$; (dashed line) $1-(T/T_\lambda)^7$;
 ($+$) the data from Ref.~\cite{Donnelly:1998}.}
\end{figure}

Here we take a minimal model:
\begin{eqnarray}
\frac{\rho_s}{\rho} = 1 - \left(\frac{T}{T_\lambda}\right)^6, \label{eq:model}
\end{eqnarray}
where $\rho (= \rho_n + \rho_s)$ is the total density of liquid helium, 
and $\rho_n$, $\rho_s$ are the normal fluid and superfluid densities,
respectively. It should be emphasized that this model does not affect a
formula for a critical temperature.
Fig.~\ref{fig:ratio} shows the superfluid density ratio as a function of
temperature. Two functions 
$1-(T/T_\lambda)^5$ and $1-(T/T_\lambda)^7$ are 
plotted, for comparison.
Fig.~\ref{fig:ratio} shows that
Eq.~(\ref{eq:model}) and experimental values are consistent 
except near the $\lambda$-transition, at which 
the superfluid density ratio is of the form $(1-T/T_\lambda)^{2/3}$.
The form of Eq.~(\ref{eq:model}) is the same as that of the Bose-Einstein
condensation in an ideal Bose gas, $1-(T/T_c)^{3/2}$.
Although the exponent ``6'' has not yet been derived, we do not consider
this problem here.

\begin{figure}[t]
\includegraphics[width=6cm]{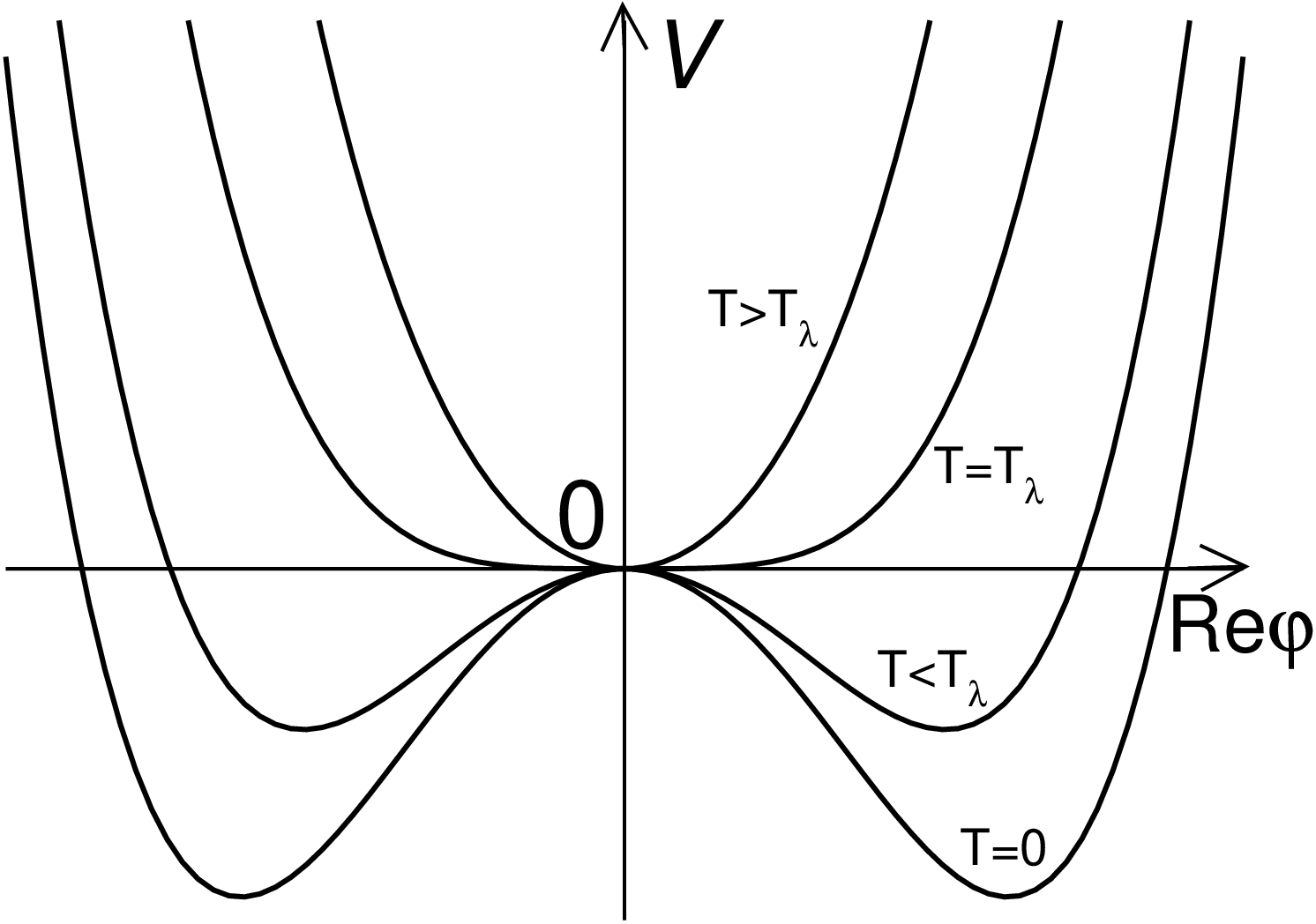}
\caption{\label{fig:transition} The schematic temperature dependence of the
 potential ${\cal V}$ for a second-order phase transition.}
\end{figure}

From Eqs.~(\ref{eq:rela}) and (\ref{eq:model}), 
we obtain the expression
\begin{eqnarray}
 \mu = \mu_0
       \left(1-\left(\frac{T}{T_\lambda}\right)^6\right), \quad
     \mu_0 = \frac{\lambda \rho_0}{m}, \label{eq:chemical}
\end{eqnarray}
where $\mu_0$ is a chemical potential at absolute zero and $\rho_0$ is
the total density at absolute zero. 
We have used the approximation $\rho \simeq \rho_0$ in
Eq.~(\ref{eq:chemical}), because the total density $\rho$ is
approximately a constant below the $\lambda$-temperature. Therefore,
Eq.~(\ref{eq:chemical}) is a good approximation at low temperature ($T <
1.7$ K). 
Fig.~\ref{fig:transition} shows
the schematic temperature dependence of the potential ${\cal V}$. 
For $T > T_\lambda$, the potential has only one minimum at $\varphi = 0$ and
the curvature at $\varphi = 0$ is positive.
When $T = T_\lambda$, the curvature at the minimum is zero, $\partial^2
{\cal V}/\partial \varphi^* \partial \varphi |_{\varphi = 0} = 0$.
For $T < T_\lambda$, the curvature at $\varphi = 0$ is negative and 
the scalar field rolls down to the $|\varphi| \not= 0$ minimum.
Thus, there is no barrier of potential between the minimum at $\varphi = 0$ and
$|\varphi| \not= 0$, indicating a second-order phase transition.

\noindent{\bf 6.}
We now derive a formula for a critical temperature of spontaneous
symmetry breaking.
When $T=0$, the field $\varphi$ is located at the nonzero minimum
$|\varphi| \not= 0$. Hence 
the global U(1) symmetry is broken in this phase.
As the temperature increases above a critical temperature, 
the potential has only one minimum at $\varphi = 0$
and the global U(1) symmetry is recovered.
From these observations, we reach a following conclusion;
{\it the depth of the potential well at absolute zero 
determines a critical temperature}.
When $T = 0$, the depth of the potential well is given by
\begin{eqnarray}
 -{\cal V}(|\varphi|=\sqrt{\mu_0/\lambda}) = \frac{\mu_0 n_0}{2}, 
\end{eqnarray}
where $n_0 = \rho_0/m$ is the number density of Bose liquid  at 
absolute zero.
This quantity $\mu_0 n_0 /2$ represents the required energy density to
recover the global U(1) symmetry of the system.
Therefore, the corresponding thermal energy density $k_B T_c n_0$
is equivalent to $\mu_0 n_0/2$. Consequently, we obtain the following relation 
\begin{eqnarray}
 k_B T_c = \frac{\mu_0}{2}.
\end{eqnarray}

\begin{figure}[t]
\includegraphics[width=5.1cm]{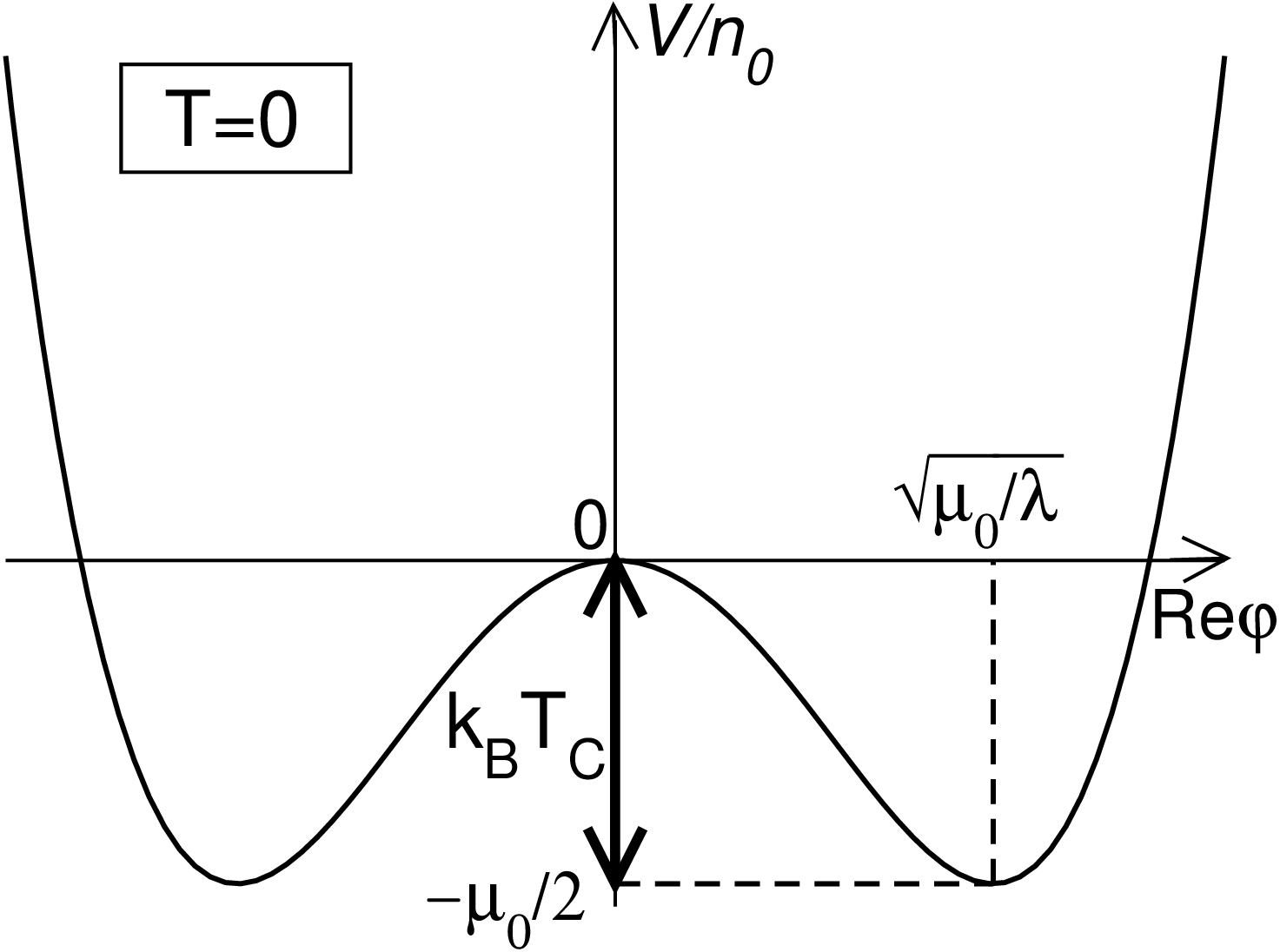}
\caption{\label{fig:zero} The absolute zero potential per number density.}
\end{figure}

Fig.~\ref{fig:zero} shows the present situation.
From above arguments, we have the following relations:
\begin{eqnarray}
 k_B T_c = \frac{\mu_0}{2}, \quad
 \frac{\rho_0}{m} = \frac{\mu_0}{\lambda}, \quad
 \lambda = \frac{2\pi \hbar^2 \sigma}{m}.
\end{eqnarray}
Therefore, a critical temperature of spontaneous symmetry breaking in
a repulsive interacting Bose system is:
\begin{eqnarray}
 T_c = \frac{\rho_0 \pi \hbar^2 \sigma}{m^2 k_B}, \label{eq:temp}
\end{eqnarray}
where $\rho_0$ is the density of Bose liquid at absolute zero, $m$ and
$\sigma$ are the mass and the hard sphere diameter of a boson, respectively.
Substituting the values of liquid helium-4 into Eq.~(\ref{eq:temp}),
$\rho_0 = 0.1451$ g/cm$^3$~\cite{Donnelly:1998},
$m = 6.6465 \times 10^{-24}$ g, and $\sigma = 2.639$
\AA~\cite{Hurly:2000,Janzen:1997}, we obtain the value $T_c = 2.194$ K.
This prediction is significantly close to the experimental value of
the $\lambda$-temperature 2.1768 K~\cite{Donnelly:1998}.
The deviation $\Delta$ between the predicted and experimental values of the
$\lambda$-temperature is less than $1\%$:
\begin{eqnarray}
 \Delta \equiv 100 \times \frac{T_c^{(theory)} -
  T_\lambda^{(exp)}}{T_\lambda^{(exp)}} 
 \simeq 0.8  \% < 1  \%.    \label{eq:Delta}
\end{eqnarray}

Thus, our prediction is much closer to the experimental value than
London's prediction.
It is because in the present approach 
the repulsive interactions between helium atoms are included in the
derivation of a formula.
While the Bose-Einstein temperature $T_{BE}$ in an
ideal Bose gas is determined by the {\it two} physical parameters, our
formula for a critical temperature, Eq.~(\ref{eq:temp}), is determined by
the {\it three} physical parameters: the density of superfluid ($\rho_0$),
the mass ($m$) and the hard sphere diameter ($\sigma$) of a boson.
The additional physical parameter $\sigma$ includes information about
interactions, and it has significantly improved the theoretical
prediction for a critical temperature.
The results are summarized in Table~\ref{tab:sum}.

\begin{table}[b]
\caption{\label{tab:sum} A comparison between BEC and SSB. Right row is
 given in this work. The deviation $\Delta$ is defined by Eq.~(\ref{eq:Delta}).}
\begin{ruledtabular}
\begin{tabular}{l|cc}
System & ideal Bose gas & interacting Bose liquid \\
Key concept  & BEC & SSB \\
$T_c$ & $2\pi\hbar^2 (n/\zeta (3/2))^{2/3}/m k_B$ 
      & $\rho_0 \pi \hbar^2 \sigma/m^2 k_B$ \\
parameter$\#$ & 2  & 3  \\
$T_c$ ($^4$He) & 3.13 K & 2.194 K \\
$\Delta$ & 44 \% & 0.8 \% 
\end{tabular}
\end{ruledtabular}
\end{table}

\begin{figure}[t]
\includegraphics[width=7cm]{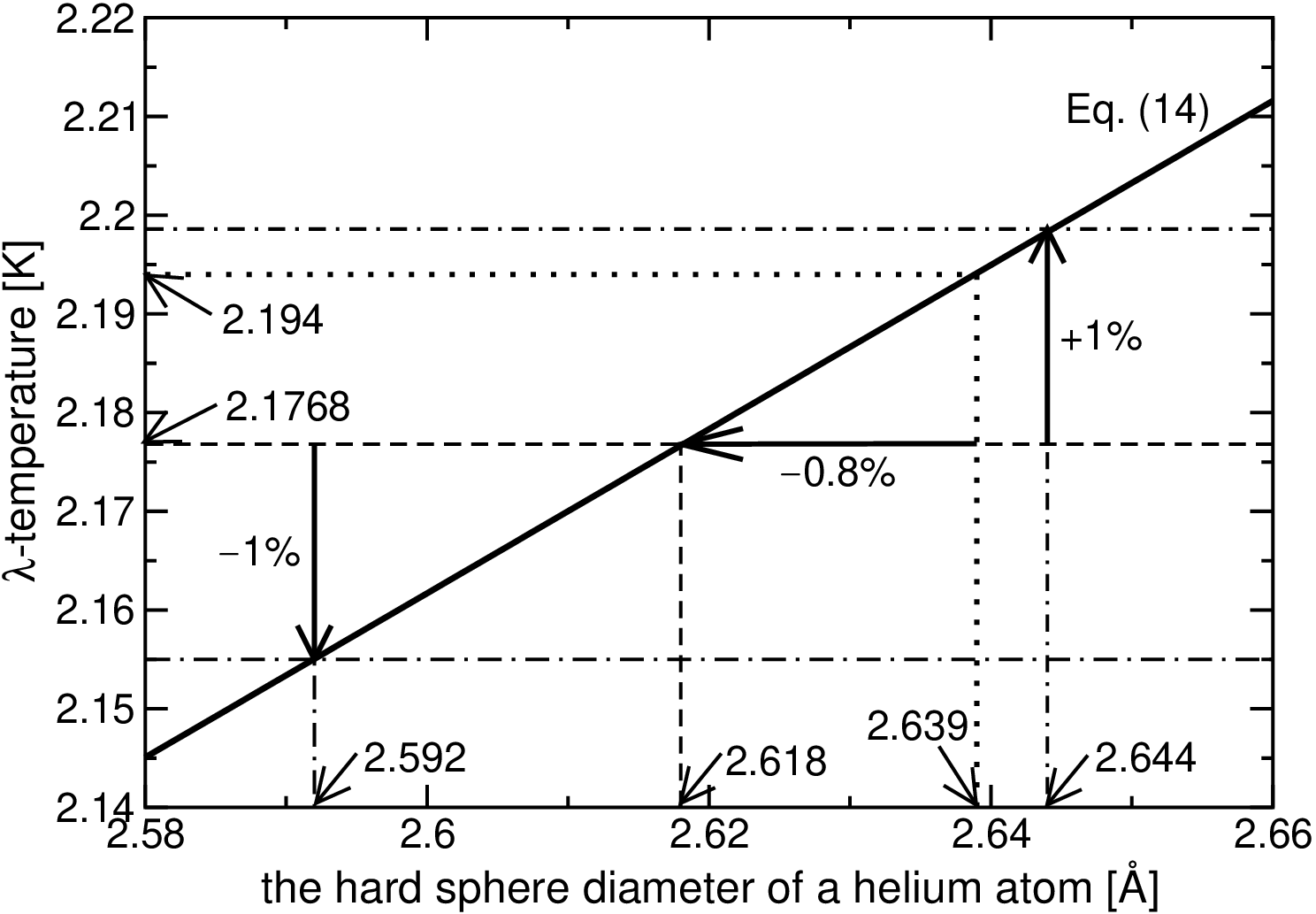}
\caption{\label{fig:diameter} The $\lambda$-temperature as a function
 of the hard sphere diameter $\sigma$ of a helium atom.}
\end{figure}

Finally, we comment on the hard sphere diameter $\sigma$.
Although $\rho_0$ and $m$ have been determined precisely, 
the uncertainty of $\sigma$ is relatively large.
The hard sphere diameter $\sigma$ represents the point at which the
interatomic potential $V(r)$ is zero ($V(\sigma) = 0$), 
and the distance $r_m$ represents the point at which the potential
$V(r)$ is minimum ($\partial V(r)/\partial r |_{r=r_m} = 0$).
Although the value of $r_m$ has been reported by many
groups~\cite{Janzen:1995}, that of $\sigma$ has been
 less reported. 
Furthermore, 
the relation $r_m = 2^{1/6} \sigma$ of the (12,6) Lennard-Jones
potential model is not applicable to the present study, because 
it is not sufficiently a good approximation.
For these reasons, we plot the $\lambda$-temperature as a function
of $\sigma$.
Fig.~\ref{fig:diameter} shows that the deviation between the predicted and 
experimental values is less than $\pm 1 \%$ for the region $2.592$ \AA
$\ \leq \sigma \leq 2.644$ \AA.

\noindent{\bf 7.}
In conclusion, 
we have derived an analytical formula for a critical temperature of
spontaneous symmetry breaking in a repulsive interacting Bose system.
The formula that we have derived, Eq.~(\ref{eq:temp}),  is a simple
analytical expression to predict $T_\lambda \simeq 2.2$ K.
We hope that this work contributes to progress of low temperature physics.



\begin{thebibliography}{99}
\bibitem{Laesecke:2002}
 A.~Laesecke,
 ``Through measurement to knowledge: the inaugural lecture of Heike
	Kamerlingh Onnes (1882),''
 J. Res. Nath. Inst. Stand. Technol. {\bf 107}(3), 261 (2002).
\bibitem{London:1938_1}
 F.~London,
 ``The $\lambda$-phenomenon of liquid helium and the Bose-Einstein 
 degeneracy,''
 Nature, {\bf 141}, 643 (1938).
\bibitem{London:1938_2}
 F.~London, 
 ``On the Bose-Einstein condensation,''
 Phys. Rev. {\bf 54}, 947 (1938).
\bibitem{Williams:1995}
 G.~A.~Williams,
 ``Specific Heat and Superfluid Density of Bulk and Confined $^4$He Near
	the $\lambda$-transition,''
 J. Low. Temp. Phys. {\bf 101}, 421 (1995).
\bibitem{Donnelly:1998}
 R.~J.~Donnelly and C.~F.~Barenghi,
 ``The Observed Properties of Liquid Helium at the Saturated Vapor Pressure,''
 J. Phys. Chem. Ref. Data. Vol. 27, No. 6 (1998).
\bibitem{Nambu:1960tm}
  Y.~Nambu,
  ``Quasi-particles and gauge invariance in the theory of  superconductivity,''
  Phys.\ Rev.\  {\bf 117} (1960) 648.
\bibitem{Arkani-Hamed:2003uy}
  N.~Arkani-Hamed, H.~C.~Cheng, M.~A.~Luty and S.~Mukohyama,
  ``Ghost condensation and a consistent infrared modification of gravity,''
  JHEP {\bf 0405}, 074 (2004).
\bibitem{Hurly:2000}
 J.~J.~Hurly and M.~R.~Moldover,
 ``Ab Initio Values of the Thermophysical Properties of Helium as
	Standards,''
 J. Res. Nath. Inst. Stand. Technol. {\bf 105}(5), 667 (2000).
\bibitem{Janzen:1997}
 A.~R.~Janzen and R.~A.~Aziz,
 ``An accurate potential energy curve for helium based on {\it ab
	initio} calculations''
 J. Chem. Phys. {\bf 107}, 914 (1997).
\bibitem{Janzen:1995}
 A.~R.~Janzen and R.~A.~Aziz,
 ``Modern He-He potentials: Another look at binding energy, effective
	range theory, retardation and Efimov states''
 J. Chem. Phys. {\bf 103}, 9626 (1995).
\end{thebibliography}
\end{document}